\documentclass{article}
\usepackage{authblk}
\usepackage[margin=1in]{geometry}
\usepackage[onehalfspacing]{setspace}
\usepackage{amsmath,mathtools}
\usepackage{pgfplots}
\usepackage{subcaption}
\usepackage{booktabs}
\usepackage[braket, qm]{qcircuit}
\usepackage{graphicx}
\usepackage{authblk}
\usepackage{xcolor}
\usepackage[linesnumbered,ruled,vlined]{algorithm2e}

\SetCommentSty{mycommfont}

\SetKwInput{KwInput}{Input}                
\SetKwInput{KwOutput}{Output}              
\SetKwProg{Fn}{Function}{}{}

\newcommand{\QWHT}{\mathcal{H}_Q}
\newcommand{\wal}{Walsh-Hadamard transform }

\usepackage[english]{babel}

\usepackage{pdflscape}
\usepackage{fancyvrb}
\usepackage{calc}
\usepackage{rotating}
\usepackage{pgf,tikz}
\usepackage{mathrsfs}
\usetikzlibrary{arrows}
\usepackage{mathtools}

\usepackage{listings}
\usepackage[]{qcircuit}
\usepackage{braket}
\usepackage{mathtools}
\usepackage{varwidth}
\usepackage{caption}
\usepackage{subcaption}
\usepackage{graphicx}
\usepackage[export]{adjustbox}

\newcommand{\norm}[1]{\left\lVert#1\right\rVert}

\makeatletter
\def\paragraph{\@startsection{paragraph}{4}%
	\z@\z@{-\fontdimen2\font}%
	{\normalfont\bfseries}}
\makeatother

\usepackage{graphicx}
\usepackage{pgffor}
\usepackage{tikz,tikz-cd}
\usetikzlibrary{backgrounds,fit,decorations.pathreplacing}  
\usepackage{booktabs}
\usepackage{enumerate}

\usepackage{amssymb, amsmath, amsthm}

\usepackage{xypic}
\usepackage{scalerel}
\usepackage{ulem}

\usepackage{graphicx}              
\usepackage{amsmath}               
\usepackage{amsfonts}              
\usepackage{amsthm}                
\usepackage{xkeyval}               
\usepackage{amssymb}
\usepackage{enumerate}             
\usepackage{color}
\usepackage{breqn}                 
\usepackage{bm}                    

\allowdisplaybreaks[1]
\usepackage{nicefrac}
\usepackage{booktabs}

\usepackage{kpfonts}
\usepackage{stackengine}
\usepackage{calc}
\newlength\shlength
\newcommand\xshlongvec[2][0]{\setlength\shlength{#1pt}%
	\stackengine{-5.6pt}{$#2$}{\smash{$\kern\shlength%
			\stackengine{7.55pt}{$\mathchar"017E$}%
			{\rule{\widthof{$#2$}}{.57pt}\kern.4pt}{O}{r}{F}{F}{L}\kern-\shlength$}}%
	{O}{c}{F}{T}{S}}

\usepackage{hyperref}
\hypersetup{
	colorlinks   = true,    
	citecolor    = red      
}

\newcommand{\RN}[1]{%
	\textup{\uppercase\expandafter{\romannumeral#1}}%
}

\allowdisplaybreaks

\newcommand{\meqref}[1]{\text{Eq}.~\eqref{#1}}

\newtheorem{thm}{Theorem}[subsection]

\newtheorem{defn}[thm]{Definition} 
 
\newtheorem{remark}[thm]{Remark}

\newcommand{\RR}{\mathbb{R}}      

\def\RR{{\mathbb R}}

\def\<{\langle}
\def\>{\rangle}

\numberwithin{equation}{section}

\begin{document}

	\title{A hybrid classical-quantum algorithm for digital image processing}
	
%
%
		
	\author[1]{Alok Shukla}
	\author[2]{Prakash Vedula}
	\affil[1]{School of Arts and Sciences, Ahmedabad University, India}
	\affil[1]{alok.shukla@ahduni.edu.in}
	\affil[2]{School of Aerospace and Mechanical Engineering, University of Oklahoma, USA}
	\affil[2]{pvedula@ou.edu}

	\maketitle

	\begin{abstract}
A hybrid classical-quantum approach for evaluation of multi-dimensional Walsh-Hadamard transforms and its applications to quantum image processing are proposed. In this approach, multidimensional Walsh-Hadamard transforms are obtained using quantum Hadamard gates (along with state-preparation, shifting, scaling and measurement operations).  
The proposed approach for evaluation of multidimensional Walsh-Hadamard transform has a considerably lower computational complexity (involving $O(N^d)$ operations) in contrast to classical Fast Walsh-Hadamard transform (involving $O(N^d~\log_2 N^d)$ operations), where $d$ and $N$ denote the number of dimensions and degrees of freedom along each dimension. 
Unlike many other quantum image representation and quantum image processing frameworks, our proposed approach makes efficient use of qubits, where only $\log_2 N $ qubits are sufficient for sequential processing of an image of $ N \times N $ pixels. 
Selected applications of the proposed approach (for $ d=2 $) are demonstrated via computational examples relevant to basic image filtering and periodic banding noise removal and the results were found to be satisfactory.
	\end{abstract}
%

	
	\section{Introduction}\label{sec:intro}
		The benefits of quantum computing algorithms over their classical counterparts have been demonstrated in the literature in several areas, including those relevant to factorization of large integers (via Shor's algorithm~\cite{shor1994algorithms, shor1999polynomial}), searching of an unstructured database for the marked entry (via Grover's algorithm ~\cite{grover1996fast, grover1997quantum} based on amplitude amplification), solutions of systems of linear equations~\cite{harrow2009quantum} and systems of linear ordinary differential equations~\cite{berry2014high}, solution of graph theory problems (based on a hybrid classical quantum algorithm known as the Quantum Approximate Optimization Algorithm~\cite{farhi2014quantum}), solution of nonlinear differential equations~\cite{shukla2021hybrid} and machine learning~\cite{wittek2014quantum}. 

		Quantum algorithms have also been shown to be offer considerable advantages in certain aspects of image processing (e.g. edge detection, pattern matching and recognition)~\cite{Wang2021, Yao2017, Ruan2021, yuan2017quantum}. A number of quantum image representations have been proposed and surveyed in Ref.~\cite{Yan2016}. 
		Some commonly considered quantum image representation formats include qudit lattice, FQRI, NEQR, QPIE or Real Ket~\cite{Ruan2016, Ruan2021, Yao2017}. 
		In the following, we will compare a few key features of our proposed approach with some of the existing quantum image processing frameworks.
		
		We note that for representation of a classical image of $N \times N$ pixels, the Real Ket format has a high storage efficiency as it just needs $2 \log_2 N$ qubits to represent the image, while FQRI and qudit lattice require $N^2$ and $2 \log_2 N +1$ qubits respectively. For instance, in the original Real Ket representation, the probability amplitudes correspond to normalized gray level values, where the normalization factor is constructed based on all pixel values in the image. 
			In our proposed work, we use a new hybrid classical-quantum approach for image processing. The image is represented classically and only $\log_2 N $ qubits are sufficient at any time for row/column based sequential processing of an image of $ N \times N $ pixels (see Remark \ref{remark_Nqubit}). 
		 Here probability amplitudes correspond to normalized gray level values of each row (or column), where the normalization factor is constructed according the pixel values in each row (or column) of the image. The requirement of fewer qubits ($\log_2 N$) in the proposed approach makes it more efficient than the original Real Ket representation (using $2 \log_2 N$ qubits)  and many other existing quantum image representation and processing frameworks \cite{Ruan2021}), especially in view of the limitations of capacity (i.e. number of qubits available), cost and measurement challenges (associated with handling of multiple qubits). The proposed representation will also be useful in the context of parallel computing (where each processor is tasked with handling rows (or columns) of data).

		For image filtering applications, we consider image representations in Walsh-Hadamard basis functions. Our choice of basis functions is primarily guided by the natural connection between the Walsh-Hadamard transform~\cite{beauchamp1975walsh} and the Hadamard gate~\cite{nielsen2002quantum} that is widely used in many quantum circuits and algorithms. Further we note that Walsh-Hadamard basis functions are already being used in classical image processing applications \cite{kuklinski1983fast}. 
		Such basis functions and associated transforms have also been found to be useful in several other fields, for example, signal processing~\cite{zarowski1985spectral}, solution of non-linear differential equations~\cite{beer1981walsh,ahner1988walsh}, solution of partial differential equations relevant to fluid dynamics ~\cite{gnoffo2014global, gnoffo2015unsteady, gnoffo2017solutions}, solution of variational problems \cite{chen1975walsh}, and cryptography~\cite{lu2016walsh}.

		The authors recently proposed a hybrid classical-quantum approach for obtaining Walsh-Hadamard transforms of arbitrary vectors~\cite{shukla2021hybrid} and applied it to solution of nonlinear ordinary differential equations. In particular, in Ref.~\cite{shukla2021hybrid}, the authors showed how useful classical information could be extracted using quantum Hadamard gates and the approach presented therein. Note that this approach primarily involves a combination of state preparation, shifting, scaling and measurement operations. It was shown that the hybrid classical-quantum approach for Walsh-Hadamard transform proposed in Ref.~\cite{shukla2021hybrid} has significantly lower computational complexity ($ \mathcal{O}(N) $ operations) in comparison to the classical Fast Walsh-Hadamard transform ($ \mathcal{O}(N \log_2 N) $ operations).
		
			While the authors' previous work~\cite{shukla2021hybrid} demonstrated a hybrid classical-quantum approach for evaluation of one-dimensional Walsh-Hadamard transforms, the present work is focused on extensions to evaluation of multi-dimensional Walsh-Hadamard transforms (using a hybrid classical-quantum approach) and applications to image processing. In the two-dimensional case, relevant to image processing, the proposed hybrid approach has a computational complexity of $O\left( N_{1} N_{2} \right)$ for computing the two-dimensional Walsh-Hadamard transform of an $ N_1 \times N_2 $ matrix (where $ N_1 $ and $ N_2 $ are powers of $ 2 $) containing image data.  In contrast, for obtaining the same result, the classical two-dimensional  fast Walsh-Hadamard transform has a computational complexity of $O\left( N_{1} ~N_{2} \log_2 (N_{1} ~N_{2} ) \right) $. Applications of the proposed hybrid approach for two-dimensional Walsh-Hadamard transforms are demonstrated through computational examples relevant to image filtering.
		More generally, we show that the computational complexity of the proposed approach for the computation of a  multi-dimensional Walsh-Hadamard transform, as defined in \meqref{eq_multi_D_Walsh_transform},
		 is  much lower  with $\mathcal{O} \left(\prod_{k=1}^{d} N_k\right)  $ operations compared to the classical Fast Walsh-Hadamard transform that needs $\mathcal{O} \left( \prod_{k=1}^{d} N_k \log_2 (\prod_{k=1}^{d} N_k) \right) $  operations.

		The rest of this paper is organized as follows. 
		In section~\ref{sec:wh_transform}, we present some basic definitions and properties of the classical Walsh-Hadamard transform and its quantum analog (implemented via the use of quantum Hadamard gates). A discussion of a hybrid classical-quantum approach for a one-dimensional Walsh-Hadamard transform is included in section~\ref{sec:hybrid-cq-wht}. Proposed extension of this hybrid approach to two-dimensional (and multi-dimensional) Walsh-Hadamard transforms, is presented in section~\ref{sec:hybrid_cq_algos_image_processing}. In section~\ref{sec:computational_examples}, applications of the proposed hybrid classical-quantum approach for two-dimensional Walsh-Hadamard transforms, in the context of image processing, are demonstrated using computational examples relevant to basic image filtering and  periodic banding noise removal.
		 Conclusions are summarized in section~\ref{sec:conclusion}.

		\section{Walsh-Hadamard Transform}\label{sec:wh_transform}
		
			Throughout this paper we will use the convention  that  $ x_i \in \{0,1\}$ will denote the $ i $-th bit in the binary representation of the integer $ x  $. More explicitly, it means, if $ x = x_0 + x_1 2 + x_2 2^2 + \ldots + x_i 2^i + \ldots + x_{n-1} 2^{n-1}$, then $ x_i $ is the $ i $-th bit in the binary representation of $ x $.

		\begin{defn}
			Let $ {\bf{v}} = [f(0) \quad  f(1) \quad  f(2) \quad  \ldots \quad f(N-1) ]^T$  be a vector with  $ N = 2^n $ components. Then its Walsh-Hadamard transform is the vector 	$ {\bf{\widehat{v}}} = [\widehat{f} (0)\quad  \widehat{f} (1) \quad \widehat{f} (2) \quad \ldots \quad \widehat{f} ({N-1})]^T  $, where  for $ k=0$, $ 1 $, $ 2 $, $ \ldots $, $ N-1 $, the component $ \widehat{f} (k)  $ of $\, {\bf{\widehat{v}}} $ is defined by 
				\begin{equation} \label{eq_def_walsh_transform}
				\widehat{f} (k) =  \frac{1}{\sqrt{N}}  \sum_{m = 0}^{N -1} \, f(m) (-1)^{\sum_{i=0}^{n-1} \, m_i k_i }.
			\end{equation}
		Here,  $ k_i $ and $ m_i $ are the $ i $-th bits in the binary representations of $ k $ and $ m $, respectively.
		Similarly, given $ {\bf{\widehat{v}}} $, its inverse Walsh-Hadamard transform is $ {\bf{v}} $ such that the component $ f(m) $ of $ {\bf{v}} $ is defined by 	\begin{equation}\label{eq_inverse_walsh_tansform}
 	f(m) =   \frac{1}{\sqrt{N}}  \sum_{k = 0}^{N -1} \, \widehat{f}(k)  (-1)^{\sum_{i=0}^{n-1} \, m_i k_i }.
 \end{equation}
		\end{defn} 
		We will write \[
		[f(0) \quad f(1) \quad f(2) \quad \ldots \quad f(N-1) ]^T  \longleftrightarrow [\widehat{f} (0) \quad  \widehat{f} (1) \quad \widehat{f} (2) \quad \ldots \quad \widehat{f} ({N-1}) ]^T 
		\]
		to denote the pair of a vector and its Walsh-Hadamard transform. 

		Alternatively, for any integer $ N=2^n $, the Walsh-Hadamard transform of the given vector $$ {\bf{v}} = [f(0) \quad  f(1) \quad \ldots \quad  f(N-1) ]^T $$ may be obtained by computing
		\begin{equation}\label{eq_hadamard_transform}
			(H^ {\otimes n}   \, \bf{v}),
		\end{equation}
	where 
	\begin{equation}\label{eq_def_H_matrix}
		H = \frac{1}{\sqrt{2}} \begin{pmatrix*}[r]
						1 & 1	\\
						1 & -1	 
						\end{pmatrix*}.
	\end{equation}
		
		A naive approach to compute the Walsh-Hadamard transform involving  the above matrix-vector multiplication is of the order $ \mathcal{O}(N^2) $ where $ N=2^n $. 
		Although, a faster classical algorithm, namely the Fast Walsh-Hadamard Transform~\cite{geadah1977natural},~\cite{beauchamp1975walsh} exists. The classical Fast Walsh-Hadamard Transform  algorithm has the time complexity of the order of $ \mathcal{O} (N \log_2 N) $ for computing the Walsh-Hadamard transform of an input vector of size $ N=2^n $.

		\subsection{Quantum Walsh-Hadamard Transform}\label{subsec:quantum_wh_transform}
		We note that the matrix $ H $ in $ \meqref{eq_hadamard_transform} $ is the transformation matrix of the quantum Hadamard gate in a computational basis. Let  $ N=2^n $ and $ {\bf{v}} = [f(0) \quad  f(1) \quad  f(2) \quad  \ldots \quad f(N-1) ]^T$  be a normalized vector, i.e., $ \norm{\bf{v}} =1 $, or equivalently, $ \sum_{k=0}^{N-1} \, (f(k))^2 = 1 $.
		We note that the quantum implementation of Walsh-Hadamard transform of $ {\bf{v}} $ involves preparing the initial state  $ \sum_{k=0}^{N-1}\, f(k) \ket{k} $, and then applying quantum Hadamard gates $ H^{\otimes n} $ on it.
		It can be verified that,
		\[
		H^{\otimes n} \left[  \, \sum_{k=0}^{N-1}\, f(k) \ket{k} \right]   = \frac{1}{\sqrt{N}} \sum_{k=0}^{N-1}\,  \left(\sum_{m = 0}^{N -1} \, f(m) (-1)^{\sum_{i=0}^{n-1} \, m_i k_i }\right) \,\ket{k} =   \sum_{k = 0}^{N -1} \, \widehat{f} (k) \,\ket{k}. 
		\]

		\section{Hybrid Classical-Quantum Approach for Walsh-Hadamard Transform}\label{sec:hybrid-cq-wht}

		A thoughtful adaptation of the quantum Walsh-Hadamard transform is central to our approach to image processing applications. 
		The Hadamard gate is one of the most useful quantum gates. One can find the Walsh-Hadamard transform to be the first step in many important quantum algorithms. It was discussed earlier that the Walsh-Hadamard transform of $ {\bf{v}} = [f(0) \quad  f(1) \quad \ldots \quad  f(N-1) ]^T $ is given by 	$   (H^ {\otimes n}   \, \bf{v}) $. Assuming that  $ \norm{{\bf{v}}} =1 $, a simple quantum circuit consisting of $ n $ Hadamard gates can compute the Walsh-Hadamard  transform of an input vector $ \bf{v} $ of size $ N=2^n $. However, the difficulty in this simple approach lies in the measurement. One can only find the square of the amplitudes of the Walsh-Hadamard transform values by carrying out the measurement. As the input sequence is assumed to be real, the components of Walsh-Hadamard transformed vector $ \widehat{\bf{v}} $ would also be real. However, the  components of  $\, \widehat{\bf{v}} \, $  may be positive or negative and this sign information is lost on carrying out the measurement. 
		We note that for many image processing applications the pixel values may be non-negative. For example, for gray scale representation the pixel values may vary from $ 0 $ to $ 255 $. However, application of the Walsh-Hadamard transforms of row/column vectors may result in vectors that contain negative values. This could potentially present challenges in unambiguous measurement as described earlier. 
		
		We tackled the core problem of obtaining Walsh-Hadamard transforms with the correct sign information in reference \cite{shukla2021hybrid}, by exploiting the structure of the Walsh-Hadamard transform matrix. 
		The approach in \cite{shukla2021hybrid} depended upon a key lemma (see Lemma 4.0.1 in \cite{shukla2021hybrid}). This resulted in an algorithm of $ \mathcal{O}(N) $ (See Algorithm 1 in \cite{shukla2021hybrid}) to compute the  Walsh-Hadamard  transform of an input vector of size $ N $. We reproduce Algorithm 1 in \cite{shukla2021hybrid} in the following for easy reference.

		\begin{algorithm}[H] \label{alg_QWHT}
			\DontPrintSemicolon
			\KwInput{The input vector $ A = [a_0 \quad a_1 \quad a_2 \quad  \ldots \quad a_{N-1} ]^{T} $ where $ N =2^n $ is a positive integer and $ a_i \in \RR $ for $ i=0 $ to $ i=N-1 $. }
			\KwOutput{The Walsh-Hadamard transform  (in the sequency order) of the input vector.}
			\Fn{$ \QWHT $ (A)}{
					$ b_0 = \epsilon + \sum_{k=0}^{N-1} \, |a_k| $ \tcp*{Here $ \epsilon $ is any positive number.}
					$   c = \sqrt{\left[ b_0^2 + \sum_{k=1}^{N-1} a_k^2 \right]}$ \tcp*{ Let $ \widetilde{A} = [b_0 \quad a_1 \quad a_2, \ldots \quad a_{N-1} ]^{T}$. Then $ c = \norm{\widetilde{A}} $. } 
					Prepare the state $ \ket{\Psi} = \frac{b_0}{c} \ket{0} + \sum_{k=1}^{N -1}\, \frac{a_k}{c} \ket{k}$ using $ n $ qubits. \tcp*{Initialize the state $ \ket{\Psi}  $ with $ \frac{\widetilde{A}}{\norm{\tilde{A}}}$.}
					Apply $ H^{\otimes } $ on $ \ket{\Psi} $. \\
					Measure all the $ n $ qubits to compute the probability $ p_k $ of obtaining the state $ \ket{k} $, for $ k=0 $ to $ 2^n-1 $. \\
					$ \delta = \frac{1}{\sqrt{N}}(b_0 - a_0 )$ \\
					$ {\bf{ u}} =   [c\sqrt{p_0} - \delta \quad c\sqrt{p_1} - \delta \quad c\sqrt{p_2} - \delta \quad \ldots \quad c\sqrt{p_{N-1}} - \delta ]^{T} $ \\
					Convert ${\bf{ u}}$ in the sequency order and store it in the vector ${ \bf{ v}} $. \\
					\Return{ the vector ${ \bf{ v}} $.}
			}
			\caption{A hybrid classical-quantum algorithm for computing the Walsh-Hadamard transform $ \QWHT(A) $ (in the sequency order) of a given input vector $ A $.}
		\end{algorithm}
		We note that the parameter $ \epsilon $ ensures that Algorithm~$ \ref{alg_QWHT} $ also works for the special case when the $ \norm{A} =0 $. 
		We already noted that the computational complexity of the classical Fast Walsh-Hadamard Transform~\cite{geadah1977natural} for an input vector of size $ N $ is of the order of $ \mathcal{O} (N \log_2 N) $, whereas our hybrid classical-quantum algorithm (Algorithm~$ \ref{alg_QWHT} $) for computation of the Walsh-Hadamard transform for an input vector of size $ N $ has a computational complexity of $ \mathcal{O}(N) $.

		\section{Hybrid Classical-Quantum Algorithms for Image Processing} \label{sec:hybrid_cq_algos_image_processing}
		Two-dimensional Walsh-Hadamard transforms will be needed for image processing applications discussed later in this work.
		However, before discussing the two-dimensional Walsh-Hadamard transforms, first we recall our the convention  that  $ x_i \in \{0,1\}$ denotes the $ i $-th bit in the binary representation of the integer $ x  $. More explicitly, it means, if $ x = x_0 + x_1 2 + x_2 2^2 + \ldots + x_i 2^i + \ldots + x_{n-1} 2^{n-1}$, then $ x_i $ is the $ i $-th bit in its binary representation. We define `bit-wise inner product' of two $ n $ bits integers $ x $ and $ y $ as 
		$ \langle x, \, y \rangle  :=  \sum_{i = 0}^{n-1} x_i y_i $. 		
		Let $ f $ be a $ N \times N $ matrix, where $ N=2^n $ for some positive integer $ n $. The two-dimensional Walsh-Hadamard transform (in the natural or Hadamard order) of the matrix $ f $ is a $ N \times N $ matrix $ F $, which may be computed as
		\begin{equation}\label{eq_two_D_Walsh_transform}
			F_{p,q} : = \frac{1}{N} \, \sum_{r=0}^{N-1} \sum_{s=0}^{N-1} \, f_{r,s} \,  (-1)^{ \langle p, \, r \rangle  + \langle q, \, s \rangle } = \frac{1}{N} \, \sum_{r=0}^{N-1} \sum_{s=0}^{N-1} \, f_{r,s} \,  (-1)^{\sum_{i=0}^{n-1}  \, p_i r_i + q_i s_i   }.
		\end{equation}
		The two-dimensional Walsh-Hadamard inverse transform (in the natural or Hadamard order) of the matrix $ F $ is given by
		\begin{equation}\label{eq_two_D_Inverse_Walsh_transform}
			f_{r,s} =  \frac{1}{N} \, \sum_{p=0}^{N-1} \sum_{q=0}^{N-1} \, F_{p,q} \,  (-1)^{\langle p, \, r \rangle  + \langle q, \, s \rangle }. 
		\end{equation}
		The normalizing constant $ \frac{1}{N} $ used in $ \meqref{eq_two_D_Walsh_transform} $ and $ \meqref{eq_two_D_Inverse_Walsh_transform} $ makes the definition of direct and inverse two-dimensional Walsh-Hadamard transforms symmetric. 
		
		One can rewrite  \meqref{eq_two_D_Walsh_transform}  as 
		\begin{equation}\label{eq_two_D_Walsh_transform}
			F_{p,q} = \frac{1}{N} \, \sum_{r=0}^{N-1}  \left(\sum_{s=0}^{N-1} \, f_{r,s} \,  (-1)^{\sum_{0}^{n-1}  \, p_i r_i } \right) \, (-1)^{\sum_{0}^{n-1}  \, q_i s_i }.
		\end{equation}
	
		It is easy to see that the  two-dimensional Walsh-Hadamard transforms of a $ N \times N $ matrix can be carried out in two steps. In the first step the  one-dimensional Walsh-Hadamard transforms of all the columns are computed (one column at a time), and then in the second step the  one-dimensional Walsh-Hadamard transforms of all the rows are computed. Similarly, The two-dimensional Walsh-Hadamard inverse transform can be implemented as a sequence of two one-dimensional Walsh-Hadamard transforms.  
		For simplicity, we discussed the two-dimensional Walsh-Hadamard transform of an $ N \times N $ matrix. Similarly, the  two-dimensional Walsh-Hadamard transforms of an $N_1 \times N_2 $ matrix (where $ N_1 $ and $ N_2 $ are powers of $ 2 $) can be carried out in two steps, by first computing the  one-dimensional Walsh-Hadamard transforms of all the columns and then computing the  one-dimensional Walsh-Hadamard transforms of all the rows. Using Algorithm~$ \ref{alg_QWHT} $, the cost for computing the one-dimensional Walsh-Hadamard transform of each column (or row) is $ \mathcal{O}(N_1) $  (or $ \mathcal{O}(N_2) $). Therefore, on taking into account the cost of computing the one-dimensional Walsh-Hadamard transforms of $ N_1 $ rows and $ N_2 $ columns, the total computational cost for computing the  two-dimensional Walsh-Hadamard transform of an $N_1 \times N_2 $ matrix (using our proposed approach, also outlined in Algorithm~$ \ref{alg_two_dimensional_Walsh_Transfrom} $) turns out to be $ \mathcal{O}(2 N_1N_2) = \mathcal{O}(N_1N_2) $. This cost is considerably less than the cost of obtaining the two-dimensional Walsh-Hadamard transforms of an $ N_1 \times N_2 $ matrix using the classical Fast Walsh-Hadamard transform 
	that has a cost of	$\mathcal{O}(N_1 N_2\log_2 N_1 N_2 ) $. More generally, it is easy to see that the cost of carrying out $ d - $dimensional Walsh-Hadamard transform as defined in \meqref{eq_multi_D_Walsh_transform},  using the proposed approach is $\mathcal{O}( \prod_{k=1}^{d} N_k)  $. In contrast, the cost of obtaining the same result using the classical Fast Walsh-Hadamard transform is  $\mathcal{O}( \prod_{k=1}^{d} N_k \log_2 (\prod_{k=1}^{d} N_k) ) $.

	    Algorithm~\ref{alg_two_dimensional_Walsh_Transfrom} computes the  two-dimensional Walsh-Hadamard transform of an $ N \times N $ matrix using the approach described above.  \\
	   	
			\begin{algorithm}[H] \label{alg_two_dimensional_Walsh_Transfrom}
			\DontPrintSemicolon
			\KwInput{A $ N \times N $ matrix $ X $. Here $ N=2^n $ for some positive integer $ n $. }
			\KwOutput{The two-dimensional Walsh-Hadamard transform of $ X $.}
			\tcc{The algorithm uses the quantum subroutine $ \QWHT $($ \bf{v} $) to compute the quantum Walsh-Hadamard transform of the input vector $ \bf{v} $ of size $ n $. }
			\Fn{$ \QWHT^{\otimes 2} $ (X)}{
				\For{$ j \gets 1$ \KwTo $ N$ }{
					$ X[j] = \QWHT (X[j]) $	\tcp*{Replace the $ j^{\text{th}} $ column of $ X $ with its Walsh-Hadamard transform.} 
				}
				\For{$ i \gets 1$ \KwTo $ N$ }{
					$ X^T[i] = \QWHT (X^T[i]) $	\tcp*{Replace the $ i^{\text{th}}$ row of $ X $ with its Walsh-Hadamard transform.} 
				}
				\Return{$ X $.}
			}
			\caption{A hybrid classical-quantum algorithm for computing two-dimensional Walsh-Hadamard transform.}
		\end{algorithm}
		
		\begin{remark} \label{remark_Nqubit}
			\leavevmode
			\begin{enumerate}[i.] 
				\item 
				In Algorithm \ref{alg_two_dimensional_Walsh_Transfrom},  the result of carrying out the one-dimensional Walsh-Hadamard transforms of a column/row is retrieved back and stored classically. The only quantum step in this algorithm involves the subroutine $ \QWHT $($ \bf{v} $) requiring $ n = \log_2 N $ qubits on a quantum computer. It means only $ n = \log_2 N $ qubits are needed for the computation of the  two-dimensional Walsh-Hadamard transforms using Algorithm \ref{alg_two_dimensional_Walsh_Transfrom}.    
				\item For computing the two-dimensional Walsh-Hadamard transforms, all the required one-dimensional Walsh-Hadamard transforms for columns (or rows) may be performed in parallel, but the total number of qubits required would be more.
\item We considered two-dimensional Walsh-Hadamard transforms above. It can easily be generalized to higher dimensions. Let $ N_k = 2^{n_k}$, $ k = 1, \, \ldots d $, then the $ d $-dimensional Walsh-Hadamard transform (in the natural or Hadamard order) 
 may be computed as
\begin{equation}\label{eq_multi_D_Walsh_transform}
	F_{p_1, \, p_2, \, \ldots,\, p_d } = \frac{1}{\left(\prod_{k=1}^{d} \sqrt{N_k}\right)}  \, \sum_{r_1=0}^{N_1-1} \sum_{r_2=0}^{N_2-1} \ldots \sum_{r_d=0}^{N_d-1} \, f_{r_1, \, r_2, \ldots,\, r_d } \,   (-1)^{ \sum_{k=1}^{d} \langle p_k, \, r_k \rangle}.
\end{equation} 
Here,   $ 0 \leq p_k, r_k  \leq N_k -1 $ and $ \langle p_k, \, r_k \rangle =  \sum_{i = 0}^{n_k-1} \,  p_{k,i} r_{k,i}  $ with $ p_{k,i}$ and $ r_{k,i} $ denoting the $ i $-th bit in the binary representation of the integer $ p_k  $ and $ r_k $, respectively.
\item As discussed earlier, the computational cost associated with obtaining the $2$-dimensional Walsh-Hadamard transform using the proposed hybrid approach (outlined in Algorithm~\ref{alg_two_dimensional_Walsh_Transfrom}) is $ \mathcal{O}(N_1N_2) $. This estimate includes classical and quantum components. The classical component including state preparation, scaling, shifting and measurement/retrieval operations has a cost of $ \mathcal{O}(N_1N_2) $. The quantum component involving the quantum Walsh-Hadamard transform has a cost of O(1). 
			\end{enumerate}
		\end{remark}

		For image processing applications, the Walsh-Hadamard transform in the sequency order is preferred because of its superior energy compaction properties in comparison to the Walsh-Hadamard transform in the natural order. We describe below quick methods of converting from one ordering of the Walsh-Hadamard transformed vector to the other.
		
		The ordering index in the natural order can be computed by reversing the bits of the gray code representation of the ordering index in the sequency order. Suppose $ s $ is the ordering index in the sequency order and one needs to obtain the corresponding ordering index $ h $ in the natural order. Then the first step is to compute the bits of the gray code $ g $ for $ s $ as follows.
		\begin{align*}
			g_{n-1} &= s_{n-1}, \\ 
			g_i &= s_i \oplus s_{i+1} \quad \text{for } i = 0, \, 1, \, 2, \, \ldots, \, n-2.
		\end{align*}
		Here, $ \oplus  $ denotes the bitwise XOR operation (or equivalently addition modulo $ 2 $). Once $ g $ is known, $ h $ is obtained by simply reversing the bits of $ g $, i.e. $ h_i = g_{n-1-i} $ for $ i = 0, \, 1, \, 2, \, \ldots, \,  n-1 $. 
		Next, suppose ordering index $ h $ in the natural order is known. Then the corresponding ordering index in the sequency order is obtained as follows. First, the bits of $ h $ are reversed to obtain $ g $. It means, $ g_i = h_{n-1-i} $ for $ i = 0, 1, 2, \ldots, n-1 $. Then $ s $ is obtained by the following computation.
		\begin{align*}
			s_{n-1} &= g_{n-1}, \\ 
			s_i &= g_i \oplus s_{i+1} \quad \text{for } i = n-2, \,  \ldots, \, 2, \,1, \, 0.
		\end{align*}

		\section{Computational Examples}\label{sec:computational_examples}
		
		In this section, we will give computational examples to illustrate the application of the two-dimensional Walsh-Hadamard transforms in image processing using our hybrid classical-quantum approach. The computation of the two-dimensional Walsh-Hadamard transforms using our our hybrid classical-quantum approach yields computational advantages 
		in comparison to a purely classical algorithm. Therefore, although the examples considered in this section can be implemented using a purely classical algorithm, they would be less efficient than our proposed method (see Sec.~\ref{sec:hybrid_cq_algos_image_processing}).

		In the following first we describe an algorithm for image filtering (Algorithm \ref{alg_main_general}), which involves the suppression of high sequency components, and then provide relevant computational examples. Subsequently, we describe algorithms for removing periodic banding noise from a given image. The cases involving images with the vertical banding noise, the horizontal banding noise and the combined horizontal and vertical banding noise will be considered. We note that the proposed algorithms for image  filtering  were successfully implemented and tested using the simulated environment of Qiskit (IBM's open source quantum computing platform).  
		
		\subsection{Example 1 - Image Filtering}

		Image filtering is a very common digital image processing technique used to suppress the noise in the image and make the image smooth. Noise is undesirable as it degrades the quality of the image. The source of noise could the camera sensor itself or the noise could get introduced during the electronic transmission of the image. 
		The \wal can be used for image filtering in the sequency domain.  Suppose $ A $ is the $ N \times N $ matrix representing the gray scale input image to be filtered. 
		Then the first step in image filtering using the \wal method is to carry out the $ 2 $-dimensional \wal of $ A $ to get the transformed image $ \tilde{A} $ in the sequency domain. Next, the image is filtered in the sequency domain. A simple filtering scheme involves suppressing certain high or low sequency components as needed. More complex filters using convolution may be used at this step depending on the requirement. Finally, the image is converted back to the spatial domain by performing the $ 2 $-dimensional inverse Walsh-Hadamard transform.

          	From the above discussion, we get the following algorithm for image filtering which involves the suppression of high sequency components.

          \begin{algorithm}[H] \label{alg_main_general}
          	
          	\DontPrintSemicolon
          	\KwInput{A $ N \times N $ matrix $ X $ corresponding to the gray scale image to be filtered. Here $ N=2^n $ for some positive integer $ n $. }
          	\KwOutput{A filtered version of $ X $, with the right bottom $ r \times r $ sequency block suppressed.}
          	\tcc{The algorithm uses the quantum subroutine $ \QWHT $($ \bf{v} $) to compute the quantum Walsh-Hadamard transform of the input vector $ \bf{v} $. }
             	{ $ X = \QWHT^{\otimes 2} $($ X $)}   \tcp*{Compute the two-dimensional Walsh-Hadamard transform of $ X $ using Algorithm~\ref{alg_two_dimensional_Walsh_Transfrom}.}
          	\For{$ i \gets N-r$ \KwTo $ N$ }
          	{	\For{$ j \gets N-r$ \KwTo $ N$ }
          		{
          			$ X[i][j] =0 $	  \tcp*{Suppress the right bottom $ r \times r $ sequency block.}
          		}
          	}
          	{ $ X = \QWHT^{\otimes 2} $($ X $)} \tcp*{Compute the two-dimensional inverse Walsh-Hadamard transform of $ X $ using Algorithm~\ref{alg_two_dimensional_Walsh_Transfrom}.}
          	\Return{$ X $.}
          	\caption{A hybrid classical-quantum algorithm for image filtering.}
          \end{algorithm}

\begin{figure}[ht]	
			\begin{center}
				\begin{subfigure}[t]{.3\textwidth}
					\centering
					\includegraphics[width=\linewidth]{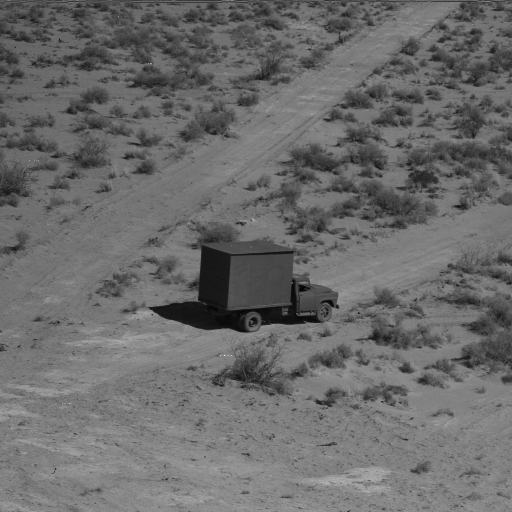}
					\subcaption{Original image, Size: $ 512 \times 512 $.}
				\end{subfigure}
				\hfill
				\begin{subfigure}[t]{.3\textwidth}
					\centering
					\includegraphics[width=\linewidth]{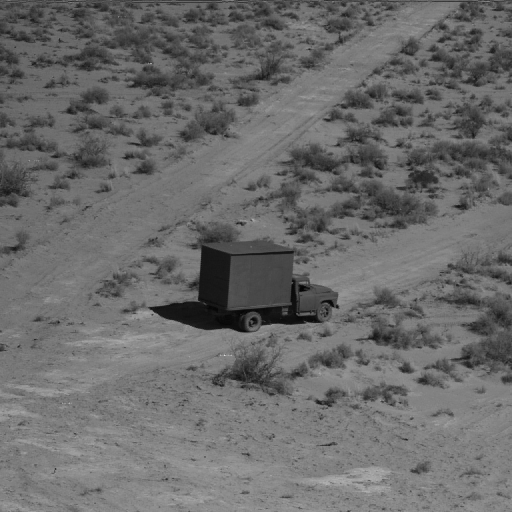}
					\subcaption{MSE: $ 4.32 $, SSIM: $ 0.98 $, PSNR: $ 41.77 $, SSB: $ 256 \times 256 $.}
				\end{subfigure} 
				\hfill
				\begin{subfigure}[t]{.3\textwidth}
					\centering
					\includegraphics[width=\linewidth]{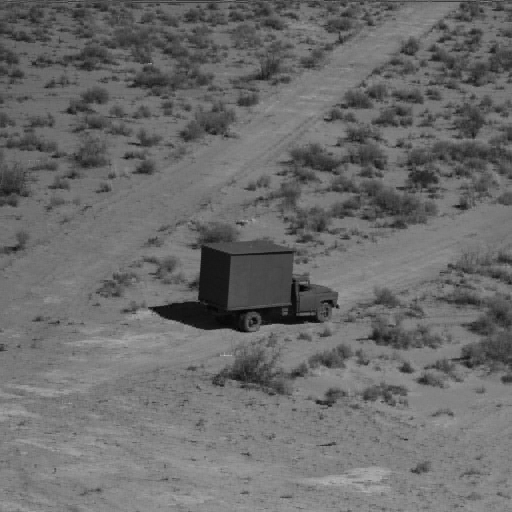}
					\subcaption{MSE: $ 18.83 $, SSIM: $ 0.91 $, PSNR: $ 35.38 $, SSB: $ 384 \times 384 $.}
				\end{subfigure} 
				\medskip \vspace{0.5cm}
				\begin{subfigure}[t]{.3 \textwidth}
					\centering
					\includegraphics[width=\linewidth]{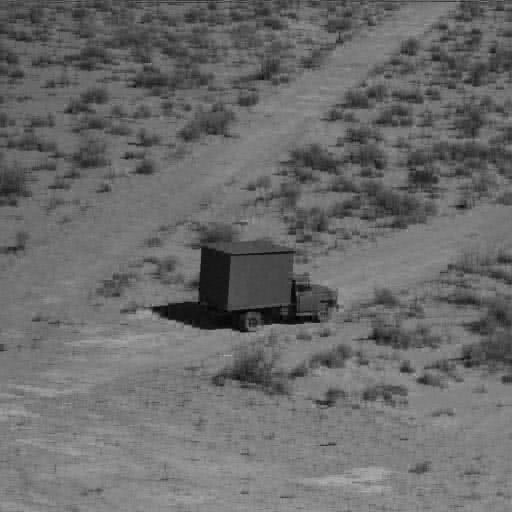}
					\subcaption{MSE: $ 48 $, SSIM: $ 0.82 $, PSNR: $ 31.32 $, SSB: $ 448 \times 448 $.}
				\end{subfigure}
				\hfill
				\begin{subfigure}[t]{.3\textwidth}
					\centering
					\includegraphics[width=\linewidth]{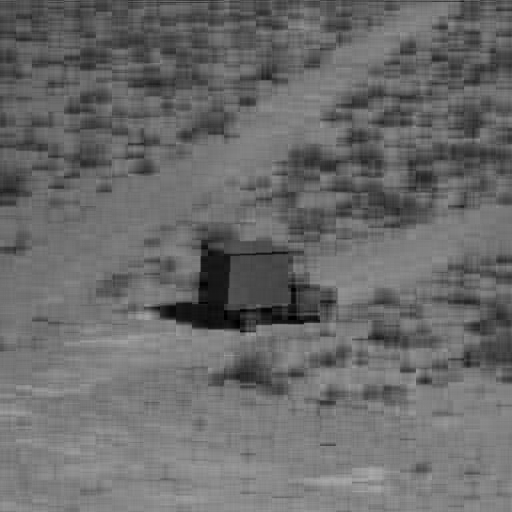}
					\subcaption{MSE: $ 96.14 $, SSIM: $ 0.73 $, PSNR: $ 28.30 $, SSB: $ 480 \times 480 $.}
				\end{subfigure}
				\hfill
				\begin{subfigure}[t]{.3\textwidth}
					\centering
					\includegraphics[width=\linewidth]{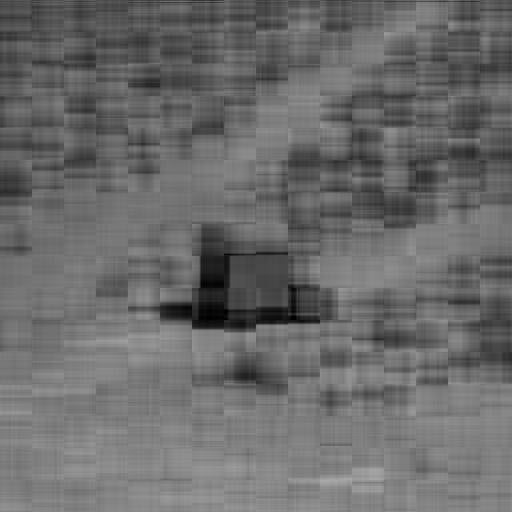}
					\subcaption{
						MSE: $ 164.13 $, SSIM: $ 0.66 $, PSNR: $ 25.98 $, SSB: $ 496 \times 496 $.}
				\end{subfigure}
			\end{center}
			\caption{Image filtering by suppressing high sequency signals. SSB denotes the size of the suppressed sequency block.}\label{fig_walsh_filter}
		\end{figure}

		The filtering of an image by suppressing the high sequency components is illustrated in Fig.~\ref{fig_walsh_filter}. The original  $n \times n $ image (where $ n =512 $) is shown in Fig.~\ref{fig_walsh_filter} (a). This image is transformed to the sequency domain by carrying out the $ 2 $-dimensional Walsh-Hadamard transform. In the sequency domain 	the elements of the submatrix $ \tilde{A}(n-r:n-r)  $ are set to $ 0 $. Here  $ \tilde{A}(n-r:n-r)  $ denotes the right bottom $ r \times r $ block of the $ n \times n $ matrix. It means all the entries in this high sequency $ r \times r $ block  is set to $ 0 $. Then by performing the $ 2 $-dimensional inverse \wal the image is converted back to the spatial domain. In Fig.~\ref{fig_walsh_filter} (b), (c), (d), (e) and (f), the size of suppressed sequency blocks are $ r \times r $  where 
		$ r = 256$, $ 384 $, $ 448 $, $ 480 $,  and $ 496 $ respectively. 
		
		Clearly, if too few sequency components are retained, then the quality of the image gets degraded.
		We recall some typical quality metrics used in digital image processing applications.  
		\begin{itemize}
			\item Mean Squared Error (MSE) between two images $ f(x,y) $ and $ g(x,y) $ of size  $ M \times N$ is defined as 
			\begin{align*}
				MSE (f,g) = \frac{1}{MN} \, \sum_{m=0}^{M} \sum_{n=0}^{N} \,  \left(f(m,n) - g(m,n)\right)^2.
			\end{align*}
			\item Peak Signal to Noise Ratio (PSNR) for two images $ f(x,y) $ and $ g(x,y) $ of size $ M \times N $ is defined as
			\begin{align*}
				PSNR (f,g) = 10 \log_{10} \left( \frac{M^2 }{ MSE (f,g)}\right),
			\end{align*}
			where $ M $ is the peak value (i.e., the maximum value) in the image data. For  grey-level (8 bits) images $ M $ is $ 255 $.
			Clearly,  as the $ 	MSE (f,g)  $ approaches zero, the $ PSNR(f,g) $ approaches infinity.
			\item Structure Similarity Index Method (SSIM) is used to measure the similarity between two images based on the  luminance,  contrast and structural correlations. This metric is very close to the human perception of similarity between two images, \cite{1284395}. 
			It is defined as 
			\begin{align*}
				SSIM (f,g) = l(x,y) \, c(x,y) \, s(x,y),
			\end{align*}
			where 
			\begin{equation*}
				\begin{aligned}
					&l(f, g)=\frac{2 \mu_{f} \mu_{g}+C_{1}}{\mu_{f}^{2}+\mu_{g}^{2}+C_{1}} \\
					&c(f, g)=\frac{2 \sigma_{f} \sigma_{g}+C_{2}}{\sigma_{f}^{2}+\sigma_{g}^{2}+C_{2}} \\
					&s(f, g)=\frac{\sigma_{f g}+C_{3}}{\sigma_{f} \sigma_{g}+C_{3}}.
				\end{aligned}
			\end{equation*}
			We note that for two images $ f $ and $ g $, the function $ l(f,g) $ captures the similarity in mean luminance ($ \mu $ denotes mean). The function $ c(f, g) $ measures the closeness in contrast ($ \sigma$ denotes standard deviation)  and the function $ s(f, g) $ gives the structural similarity between the two images $ f $ and $ g $ ($\sigma_{f g}  $ denotes covariance between $ f $ and $ g $).   
			\end{itemize}
		
			Table~$ \ref{Table} $ below describes how the quality of the filtered image depends on the  size of the suppressed block in the sequency domain.
		
		\begin{table}[ht]
			\begin{center}
				\begin{tabular}{@{}cccccc@{}}
					\toprule
			\textbf{Figure} 	&	\textbf{Size of the suppressed block} & \textbf{MSE} & \textbf{PSNR} & \textbf{SSIM} & \textbf{Image size}
					\\
					\toprule
					\toprule
							\\
		\ref{fig_walsh_filter} (a) 	&		$ 	0 \times 0 $ & $ 0  $ & $ 0 $ & $ \infty $ & $ 258 $ KB 
					\\
					\midrule
		\ref{fig_walsh_filter} (b)	&		$ 	256 \times 256 $ & $ 4.32  $ & $ 0.98 $ & $ 41.77 $ &  $ 252 $ KB
					\\
					\midrule
		\ref{fig_walsh_filter} (c)	&		$ 384 \times 384 $ & $ 18.83 $ & $ 35.38 $ &  $ 0.91 $ & $ 228 $ KB
					\\
							\midrule
		\ref{fig_walsh_filter} (d)	&		 $ 448 \times 448 $ & 	$ 48 $ &    $ 31.32 $ &  $ 0.82 $ & $ 196 $ KB
					\\
				\midrule
		\ref{fig_walsh_filter} (e)	&		$ 480 \times 480 $ &  $ 96.14 $ & $ 28.30 $ &  $ 0.73 $ &  $ 167 $ KB
			\\
				\midrule
		\ref{fig_walsh_filter} (f)	&		$ 496 \times 496 $ &  $ 164.13 $ &   $ 25.98 $ & $ 0.66 $ & $ 143 $ KB
			\\
					\midrule
					\\
					\bottomrule
					\bottomrule
				\end{tabular}
				\caption{Quality parameters of the filtered image.} \label{Table}
			\end{center}
		\end{table}

\begin{figure}[ht]	
			\begin{center}
				\begin{subfigure}[t]{.2\textwidth}
					\centering
					\includegraphics[width=\linewidth]{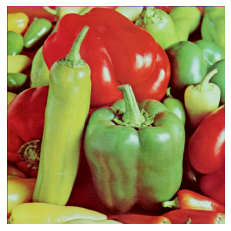}
					\subcaption{Original image, Size: $ 512 \times 512 $. } 
				\end{subfigure}
				\hspace{0.4cm}
				\begin{subfigure}[t]{.2\textwidth}
					\centering
					\includegraphics[width=\linewidth]{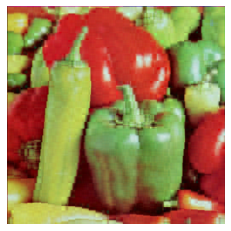}
					\subcaption{SSB: $ 480 \times 480 $.}
				\end{subfigure} 
				\hspace{0.4cm}
				\begin{subfigure}[t]{.2\textwidth}
					\centering
					\includegraphics[width=\linewidth]{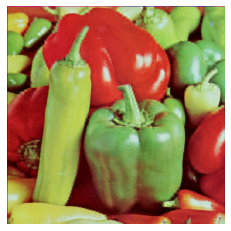}
					\subcaption{SSB: $ 448 \times 448 $.}
				\end{subfigure} 
				\hfill
			\end{center}
			\caption{Image filtering by suppressing high sequency signals for each of the RGB channels of a colored image. SSB denotes the size of the suppressed sequency block.}\label{fig_color}
		\end{figure}

		The same approach can be used for filtering of color images via the application of the above algorithm for each of the RGB (Red,  Green and Blue) channels.  
		The original image of size $ 512 \times 512 $ is shown in  Fig.~\ref{fig_color} (a). Fig.~\ref{fig_color} (b) and (c) are obtained by suppressing sequency blocks as discussed earlier. The size of suppressed sequency block is $ 480 \times 480  $ for Fig.~\ref{fig_color} (b) and  $ 448 \times 448 $ for Fig.~\ref{fig_color} (c). We note that even after suppressing  sequency block of size  $ 448 \times 448 $, with naked eye the resulting image shown Fig.~\ref{fig_color} (c) appears as good as the original image in Fig.~\ref{fig_color} (a).

		\subsection{Example 2 - Periodic Banding Noise Removal}
		As our next example, we take up the removal of periodic banding noise using the Walsh-Hadamard transform approach. Such periodic banding noise is known to occur in digital photography and image capturing, for example in satellite images associated with differences between the forward and reverse scans of the sensor \cite{quirk1992technique}. 
\subsubsection{Periodic Vertical Banding Noise}

	\begin{figure}[ht]	
			\begin{center}
			\begin{subfigure}[t]{.3\textwidth}
				\centering
				\includegraphics[width=\linewidth]{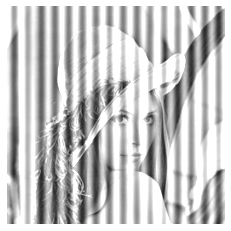}
				\subcaption{Original image with a periodic vertical banding noise.} 
			\end{subfigure}
						\hspace{0.8cm}
			\begin{subfigure}[t]{.3\textwidth}
				\centering
				\includegraphics[width=\linewidth]{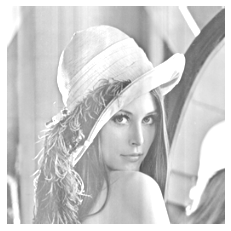}
				\subcaption{Filtered image with the periodic vertical banding noise removed.}
			\end{subfigure} 
			\hfill
		\end{center}
		\caption{Vertical banding noise removal.}\label{fig_vertical}
	\end{figure}

The image in Fig.~\ref{fig_vertical}(a) contains a vertical periodic banding noise. A hybrid classical-quantum approach for removing this  vertical periodic banding noise is described in Algorithm \ref{alg_main_noise_vertical}. We note that Algorithm \ref{alg_main_noise_vertical} begins by computing the $ 2 $-dimensional Walsh-Hadamard transform of the matrix representing the noisy image. Then the key step in Algorithm \ref{alg_main_noise_vertical} is to suppress all the elements (i.e., the sequency components) in the first row except the first element in the first row of the transformed matrix (see Step 3 in Algorithm \ref{alg_main_noise_vertical}). Finally, the  $ 2 $-dimensional inverse Walsh-Hadamard transform is carried out to get back the filtered image with the vertical periodic banding noise removed. The result of applying Algorithm \ref{alg_main_noise_vertical} to the noisy image shown in Fig.~\ref{fig_vertical}(a) results in the image in Fig.~\ref{fig_vertical}(b). It is clear that the image in Fig.~\ref{fig_vertical}(b) is mostly free of the periodic vertical banding noise present in the input image Fig.~\ref{fig_vertical}(a).

		\begin{algorithm}[H] \label{alg_main_noise_vertical}
			\DontPrintSemicolon
			\KwInput{A $ N \times N $ matrix $ X $ corresponding to the gray scale. The image contains a vertical band noise which needs to be filtered. Here $ N=2^n $ for some positive integer $ n $. }
			\KwOutput{A filtered version of $ X $ with the vertical band noise removed.}
				\tcc{The algorithm uses the quantum subroutine $ \QWHT $($ \bf{v} $) to compute the quantum Walsh-Hadamard transform of the input vector $ \bf{v} $. }
			{ $ X = \QWHT^{\otimes 2} $($ X $)}   \tcp*{Compute the two-dimensional Walsh-Hadamard transform of $ X $ using Algorithm~\ref{alg_two_dimensional_Walsh_Transfrom}.}
				\For{$ j \gets 2$ \KwTo $ N$ }
				{
					$ X[1][j] =0 $	  \tcp*{Suppress all but the first element in the first row in the sequency domain}
				}
			{ $ X = \QWHT^{\otimes 2} $($ X $)} \tcp*{Compute the two-dimensional inverse Walsh-Hadamard transform of $ X $ using Algorithm~\ref{alg_two_dimensional_Walsh_Transfrom}.}
			\Return{$ X $.}
			\caption{A hybrid classical-quantum algorithm for the removal of vertical banding noise.}
		\end{algorithm}
	
	\subsubsection{Periodic Horizontal Banding Noise}
	
		\begin{figure}[ht]	
		\begin{center}
			\begin{subfigure}[t]{.3\textwidth}
				\centering
				\includegraphics[width=\linewidth]{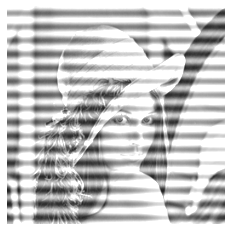}
				\subcaption{Original image with a periodic horizontal banding noise.} 
			\end{subfigure}
			\hspace{0.8cm}
			\begin{subfigure}[t]{.3\textwidth}
				\centering
				\includegraphics[width=\linewidth]{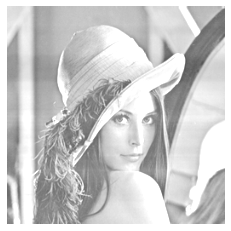}
				\subcaption{Filtered image with the periodic horizontal banding noise removed.}
			\end{subfigure} 
			\hfill
		\end{center}
		\caption{Horizontal banding noise removal.}\label{fig_horizontal}
	\end{figure}
	
Next, we consider an image,  Fig.~\ref{fig_horizontal}(a),  containing horizontal periodic banding noise. The method of removing this noise is quite similar to the approach in Algorithm \ref{alg_main_noise_vertical}. The only change needed in Algorithm \ref{alg_main_noise_vertical} is to modify Step $ 3 $ such that instead of suppressing elements in the first row (except the first element) in the sequency domain, elements in the first column (except the first element) in sequency domain are suppressed.
 The image in Fig.~\ref{fig_horizontal}(b) shows the resulting image after the removal of horizontal periodic banding noise from Fig.~\ref{fig_horizontal}(a).  
	
\subsubsection{Combined Horizontal and Vertical Banding Noise}
	
\begin{figure}[ht]	
	\begin{center}
		\begin{subfigure}[t]{.3\textwidth}
			\centering
			\includegraphics[width=\linewidth]{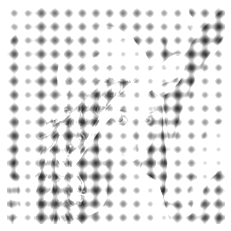}
			\subcaption{Original image with both horizontal and vertical banding noise.} 
		\end{subfigure}
		\hspace{0.8cm}	
		\begin{subfigure}[t]{.3\textwidth}
			\centering
			\includegraphics[width=\linewidth]{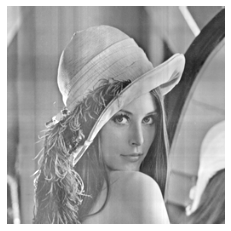}
			\subcaption{Filtered image with both the horizontal and vertical banding noise removed.}
		\end{subfigure} 
		\hfill
	\end{center}
	\caption{Both horizontal and vertical banding noise removal.}\label{fig_band}
\end{figure}

	Suppose an image contains both the horizontal and the vertical periodic banding noise as in  Fig.~\ref{fig_band}(a). By suppressing elements in the first column and also the elements in the first row (except the element at the position $ (1,1) $, i.e. the top left element) in the transformed matrix in the sequency domain, one can remove both the vertical periodic banding noise from Fig.~\ref{fig_band}(a). The method is described in detail in Algorithm \ref{alg_main_noise_band}. The `offset' value in Algorithm \ref{alg_main_noise_band} controls the intensity of the filtered image. The resulting filtered image with the `offset' value $ 150 $ in Algorithm \ref{alg_main_noise_band} is shown in Fig.~\ref{fig_band}(b).

		\begin{algorithm}[H] \label{alg_main_noise_band}
		\DontPrintSemicolon
		\KwInput{A $ N \times N $ matrix $ X $ corresponding to the gray scale. The image contains a vertical band noise which needs to be filtered. Here $ N=2^n $ for some positive integer $ n $. }
		\KwOutput{A filtered version of $ X $ with the vertical band noise removed.}
			\tcc{The algorithm uses the quantum subroutine $ \QWHT $($ \bf{v} $) to compute the quantum Walsh-Hadamard transform of the input vector $ \bf{v} $. }
		{ $ X = \QWHT^{\otimes 2} $($ X $)}   \tcp*{Compute the two-dimensional Walsh-Hadamard transform of $ X $.}
		\For{$ j \gets 2$ \KwTo $ N$ }
		{
			$ X[1][j] =0 $	  \tcp*{Suppress all but the first element in the first row in the sequency domain}
			$ X[j][1] =0 $     \tcp*{Suppress all but the first element in the first column in the sequency domain}
		}
	   $ X[1][1] = N \times $  offset  \tcp*{The `offset' value controls the intensity of the filtered image.}
		{ $ X = \QWHT^{\otimes 2} $($ X $)} \tcp*{Compute the two-dimensional inverse Walsh-Hadamard transform of $ X $.}
		\Return{$ X $.}
		\caption{A hybrid classical-quantum algorithm for the removal of combined horizontal and vertical banding noise.}
	\end{algorithm}

		\section{Conclusion}\label{sec:conclusion}

In this work, we proposed a hybrid classical-quantum approach for obtaining multi-dimensional Walsh-Hadamard transforms of arbitrary real fields with applications to quantum image processing. These multidimensional Walsh-Hadamard transforms are obtained using quantum Hadamard gates (along with state-preparation, shifting, scaling and measurement operations) and can be considered as generalization of evaluation of one-dimensional Walsh-Hadamard transforms of arbitrary vectors presented in a recent work~\cite{shukla2021hybrid}. 
Our proposed approach makes efficient use of qubits as it needs only $\log_2 N $ qubits for sequential processing of an image of $ N \times N $ pixels. This representation can be considered to more efficient, when compared to many other commonly used quantum image representations discussed in the literature. This becomes an important advantage especially considering the scarcity of available qubits in current generation of quantum computers.

We note that the computational cost of the proposed approach for the computation of a  $ d - $dimensional Walsh-Hadamard transform, 
as defined in \meqref{eq_multi_D_Walsh_transform},
is  considerably lower (with $\mathcal{O} \left(\prod_{k=1}^{d} N_k\right)  $ operations) compared to the classical Fast Walsh-Hadamard transform (with $\mathcal{O} \left( \prod_{k=1}^{d} N_k \log_2 (\prod_{k=1}^{d} N_k) \right) $  operations).
The proposed approach for obtaining the Walsh-Hadamard transform of image data (where $ d=2 $) was implemented and tested on the simulation environment on Qiskit (IBM's open source quantum computing platform). Applications of the proposed approach were successfully demonstrated via computational examples relevant to basic image filtering and periodic banding noise removal. 
Future work could involve applications of the core methodology proposed here to image, video and/or high-dimensional data compression.

%
			\bibliographystyle{unsrt}

	\end{document}